\documentclass[a4paper]{jpconf}
\usepackage{graphicx}
\begin{document}
\title{Photon measurements in forward rapidity in heavy-ion collisions}

\author{Yogendra Pathak Viyogi}

\address{Variable Energy Cyclotron Centre, 1/AF, Bidhan Nagar, Kolkata 700064}

\ead{viyogi@veccal.ernet.in}

\begin{abstract}

Contribution of a  preshower photon multiplicity detector to the physics of
ultra-relativistic nuclear collisions is reviewed and future possibilities at 
RHIC and LHC are discussed.

\end{abstract}.

\section{Introduction}
One of the main goals of the high energy heavy-ion collision 
program is to look for the possible formation of Quark-Gluon Plasma (QGP).
In the collision process the system may go through a phase transition 
from confined hadronic matter to a deconfined state of quarks and gluons 
which will be a 
transient state. The system then
 evolves from a very hot and dense QGP state to normal 
hadronic matter by undergoing cooling, expansion and hadronization. 
Information obtained from various 
global observables (e.g photon multiplicity ($N_{\mathrm \gamma}$), 
charged particle 
multiplicity ($N_{\mathrm ch}$) and transverse energy) is used to
characterize the system formed in such collisions. 
Study of photon production has shown a great promise in studying the 
various aspects of the reaction mechanism in such collisions and dynamics 
of particle production~\cite{
wa93-flow,wa93-pt,wa93-eta,
wa98-global-dcc,wa98-eta,wa98-local-dcc,wa98_fluc,wa98-central-dcc,
wa98-flow,starphoton}.

The measurement of photons in high energy physics experiments 
has been traditionally carried out using 
calorimeters. Due to large spatial density of produced particles in the
forward rapidity region in ultra-relativistic nuclear collisions, 
and consequent overlap of showers, one cannot
use calorimeters beyond a certain region. In such a situation, a limited
goal of photon study can be achieved using a preshower detector having a
relatively thinner converter and restricting the development of shower. 
The preshower detector can only measure the spatial distribution of photons
but not their energies. 

Such a preshower photon multiplicity detector (PMD) has been employed in 
 SPS and RHIC experiments and is planned also for the LHC experiment. In 
addition to extending the phase space coverage for the measurement of
pseudorapidity distributions, the PMD data on spatial distribution of photons,
along with other 
measurement of charged particle multiplicity and event centrality, has been
able to address the 
following important physics topics related to phase transition and 
chiral symmetry restoration:

\begin{itemize}

\item determination of the reaction plane and the probes of thermalization via
studies of azimuthal anisotropy and flow;

\item critical phenomena near the phase boundary
leading to fluctuations in global observables like
multiplicity  and
pseudorapidity distributions;

\item signals of chiral-symmetry restoration (e.g. disoriented chiral condensates) 
through the measurement of
charged-particle multiplicity ($ N_{\rm ch}$) also
 in a common part of phase space and study of the observables
  $N_\gamma/N_{\rm ch}$ with full azimuthal coverage.

\end{itemize}

\section{Photon Multiplicity Detector}

The preshower detector for measuring photon multiplicity consists of a thin 
converter behind which is placed
 an array of sensitive detection elements (cells or pads).
The thickness of the
converter is chosen such that it provides a reasonably large signal for
electromagnetic particles, restricts shower size to a few cells and the 
response to charged hadrons is limited to preferably one pad or cell. The 
thickness of the converter and the granularity (size, shape of one detection
element) have to be optimized using Monte-Carlo simulations. 
A second plane of similar granularity may be placed in front of the converter
for improving the discrimination between charged hadrons and photons.

\subsection{Detector for the  SPS experiments}

The PMD for the fixed target SPS experiments was
 made using plastic scintillator pads, with 
scintillation light being transported to readout devices
using wavelength shifting fibers. The 
light at the fibre ends were converted to electrical signals and readout using
the image intensifier (II) and CCD camera systems obtained from the old 
UA2 experiment at CERN.
The basic parameters of the two detectors, used in the WA93 and WA98
experiments, are summarized in Table~\ref{pmd-sps}. These  had only 
one plane of detection elements behind a 3~X$_0$ thick lead converter. 

\subsubsection{PMD for the WA93 experiment :}

The first PMD, made for the WA93 experiment,  consisted
 of 7500 plastic scintillator pads and had long WLS fibers for transporting
scintillation light to the II+CCD  camera readout system~\cite{wa93-nim}.
The pads were square in shape and having
uniform size of 2 cm. 
These were placed in four quadrants of a light-tight box,
each quadrant having 1875 pads arranged in a matrix of 50~$\times$~38 
with a space for 5~$\times$~5 pads left for the hole of the beam pipe. The 
fiber bundle from each quadrant was read out using one camera system.

The detector, mounted at 10.5m from the target, covered the
pseudorapidity region 2.8 $\le \eta \le$ 5.2 of which the region 
3.3$\le\eta\le$4.9 had full azimuthal coverage. 
It was installed in 1991 and was used to study the
collisions of 200~A.GeV sulphur ions with gold nuclei during 1991
and 1992 SPS run periods.

\subsubsection{PMD for the WA98 Experiment :}

The PMD for the WA98 experiment  was also based on the
plastic scintillator pads, with light being transported using WLS fibres~\cite{wa98-nim}.
Based on the experience gained during the operation of the WA93 PMD, the
fabrication technology was modified for 
improvements in light collection and  reduction of the distortion
of the image at the CCD. The WLS fibres of short length were coupled to long 
lengths of clear fibres to reduce light attenuation and a black paint was
used in bundling the fibre ends to reduce spreading of the image on the CCD.
To make the
hit density nearly uniform over the detector, four different pad sizes, ranging
from 15mm square to 25 mm square, were used,
 inner part
having smaller pads and 
outer parts having larger pads. 
The detector 
consisted of 28 individual light-tight 
box modules, each having a matrix of 50$\times$38 
scintillator pads read out using one image intensifier and CCD 
camera system.

\begin{table}
\begin{center}
\caption{Parameters of WA93 and WA98 PMDs}
\label{pmd-sps}
\vskip 4mm
\begin{tabular} {|l|l|l|} \hline
Parameter & WA93 & WA98 \\ \hline
($\eta,\phi$) coverage & 2.8-5.2 (2$\pi$ in 3.3-4.9) & 2.9-4.2 (2$\pi$ in 3.2-4.0) \\
Number of readout cameras & 4 & 28 \\
Number of pads & 7500 & 53200 \\
Distance from target & 10.5~m & 21.5~m \\
Pad sizes (square shape) & 20 mm  & 15 mm, 20 mm,\\ 
& & 23 mm, 25 mm \\
Pad thickness & 3 mm & 3 mm \\ 
Light transport & WLS fibres & WLS + Clear fibres \\ \hline
\end{tabular}
\end{center}
\end{table}

The detector extended to an area of 21 sq.m. It was installed at 21.5~m from 
the target  and covered the pseudo-rapidity
region 2.5 $\le \eta \le$ 4.2, of which the region $\eta$=
3.2-4.0 had full azimuthal coverage. It is shown in Fig.~\ref{wa98-pmd}.
It was installed in the 
experiment in 1994 and took data for lead ion induced collisions at 158.A~GeV energy
during 1994 to 1996 SPS run periods.  

\subsection{Detector for colliders}

For the next generation collider experiments at RHIC and LHC, it was soon 
realized that the PMD would have be made with a different technology.
Scintillator pads with fibres were bulky and the readout systems
 were quite expensive. The collider environment required sleek 
detectors which could fit in a small space. It was also realized that 
with the increased multiplicity of produced particles, the detector design
criteria had to be redefined, in particular charged particle hits on the
preshower plane had to confined to a single pad or cell. Due to the presence
of substantial number of electrons, the use of second plane in front of the
converter also became necessary.

Using gas detectors, low cost readout could be obtained  in the form of
signal processing ASICs. However, existing gas detector technologies  did not 
provide the solution to the containment of charged particle hits to a single 
detection unit.

An extensive R\&D effort resulted in the design of a cellular honeycomb 
proportional counter, using Ar(70\%)+CO$_2$(30\%) as the sensitive
medium and with novel ideas of extended cathode which provided
almost uniform response to charged particles within the detector volume.
This design offered all the advantages of lower cost and ease of
 construction for large area and highly granular detector system.
The results of such an R\&D are summarized in 
Ref.~\cite{alice-pmd-tdr,alice-nim}. 

The basic parameters of the PMD for the RHIC and LHC experiments  are 
given in Table~\ref{pmd-collider}. In these detectors also we used 3~X$_0$
thick lead converter.

\subsubsection{PMD for the STAR experiment at RHIC :}

The STAR PMD~\cite{star-nim}
 consists of a set of super-modules arranged in two vertical halves
so that each half moves independently around the beam pipe. The super-modules
are of various shapes but all are made up of rhombus unit modules having
24$\times$24 cells. The number of unit modules in the supermodules vary from 4 
to 9. There are two planes of the sensitive detectors, one in 
front of the converter (acting as charged particle veto) and one behind the 
converter which registers preshower signals. There are 12 supermodules
 in each plane of the detector. 

The signal processing for STAR PMD is done using GASSIPLEX chips, which 
has 16 channels of
preamplifier and shaper and provides analog multiplexed output.
One front-end electronics (FEE)-Board has 4 chips and is 
connected to a group
of 64 cells in a 8$\times$8 matrix on the detector. 
The digitization and readout of the analog multiplexed signals 
 is done using C-RAMS modules. The track and hold flag is generated
using a pre-trigger signal based on beam-beam counter in STAR
 because the actual Level 
0 signal arrives rather late. Each block of C-RAMS handles signals from a chain
of 27 FEE boards, i.e., for 1728 cells. There are in all 48 chains read out
using 24 C-RAMS modules.

The detector is mounted in the STAR experiment at 540cm from the center of the
TPC and nominally covers the pseudorapidity region 2.3-3.8 with full
azimuthal acceptance.

\begin{figure}[h]
\begin{minipage}{18pc}
\includegraphics[width=18pc]{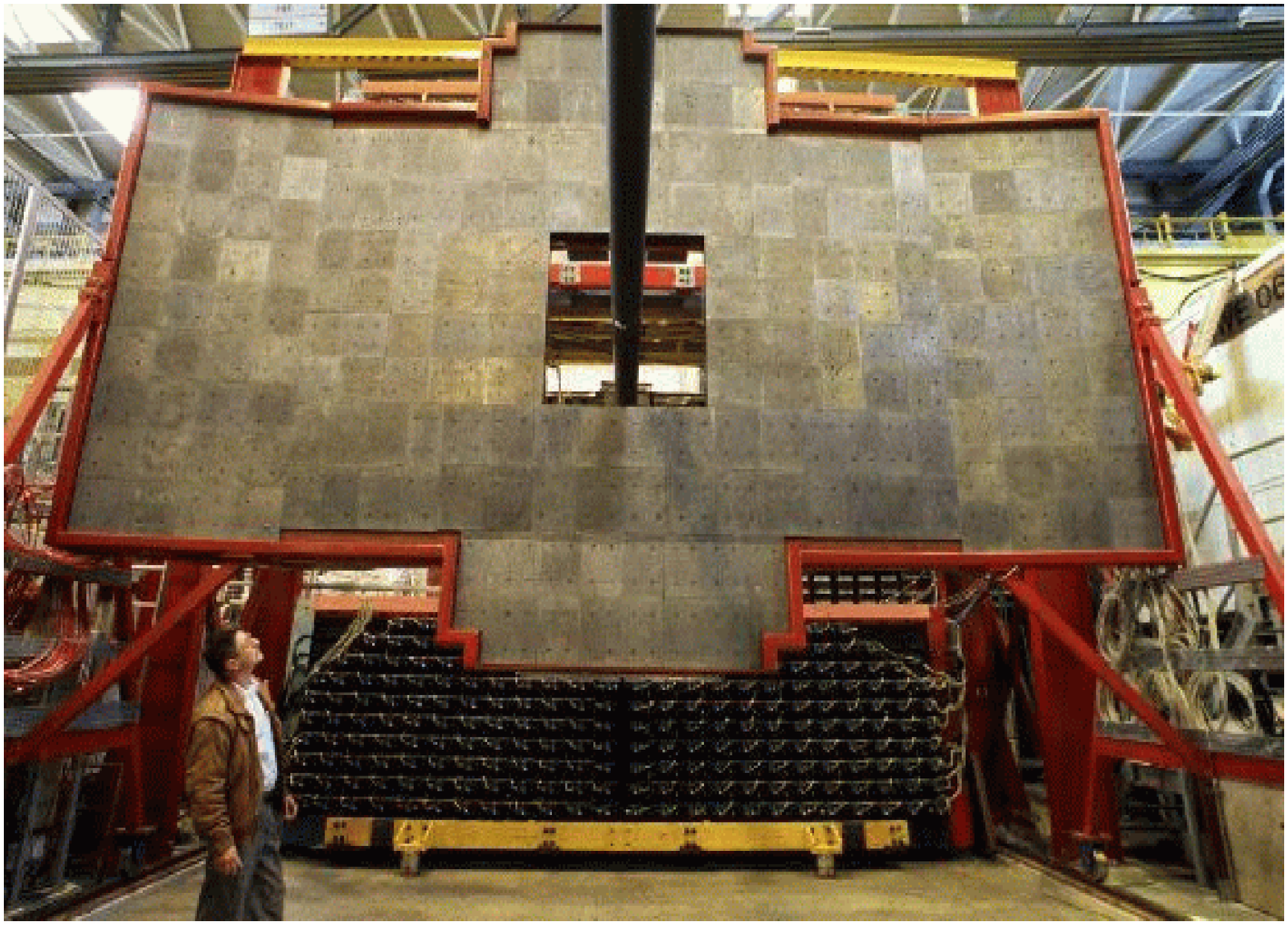}
\caption{\label{wa98-pmd}
View of the  WA98 PMD from the converter side. The sections on four sides were 
suitably inclined to provide near-normal incidence to incoming particles.}
\end{minipage}\hspace{2pc}%
\begin{minipage}{18pc}
\includegraphics[width=18pc]{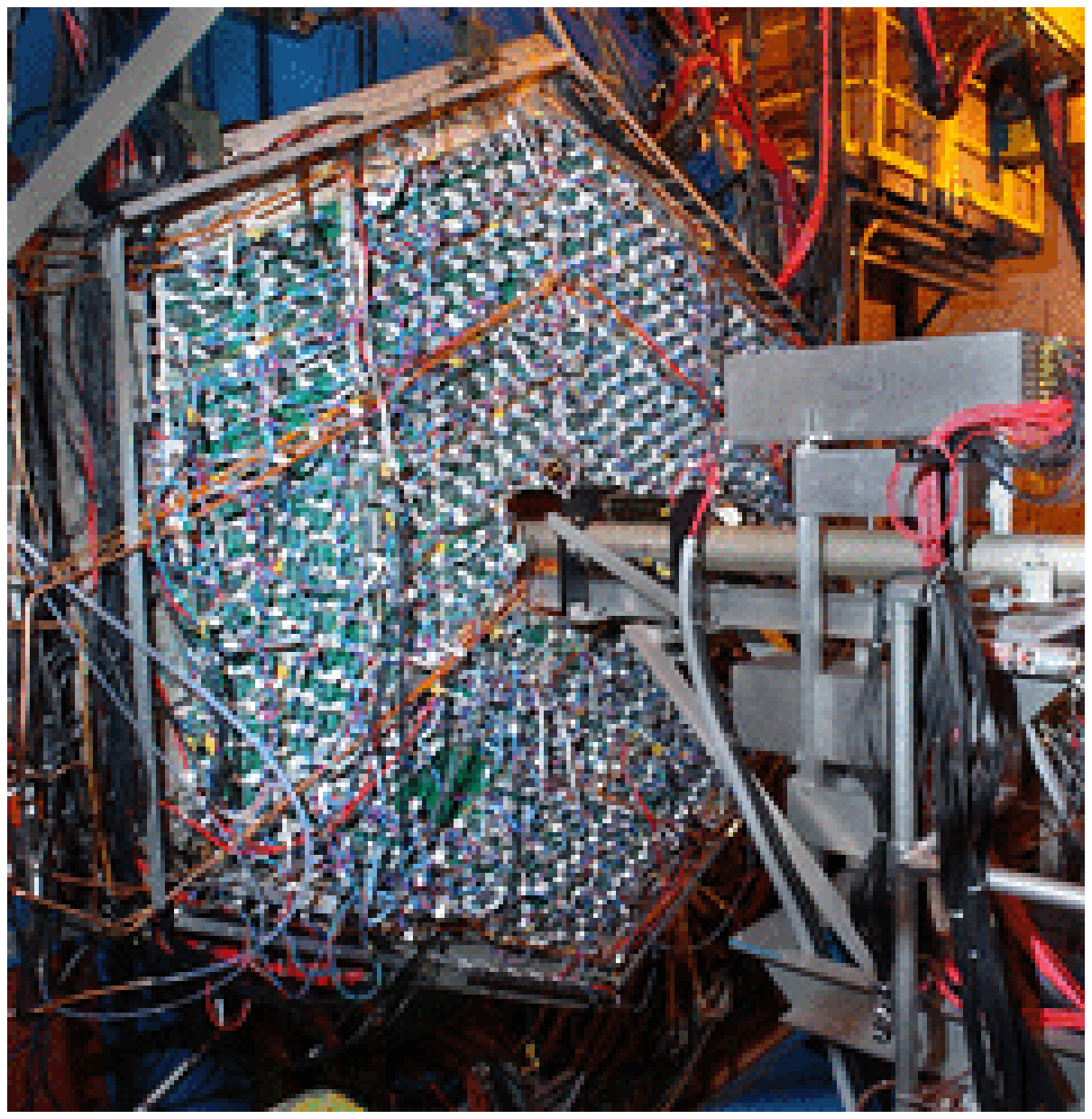}
\caption{\label{star-pmd}
View of the preshower plane of STAR PMD from the RHIC tunnel side.}
\end{minipage} 
\end{figure}

Fully instrumented preshower plane of the STAR PMD is shown in 
Fig.~\ref{star-pmd} in the data taking mode when the two halves are touching each other.

\subsubsection{PMD for the ALICE experiment at the LHC :}

The PMD for the ALICE experiment consists of hexagonal cells having 0.23 cm$^2$
cross-section and 5~mm depth. Although originally proposed to use the cells
identical to those used in the STAR PMD, the detector geometry was modified 
later
due to  constraints of
available space within the ALICE experiment ~\cite{alice-atdr,pprv1}.

The detector consists of rectangular shaped 
unit modules having 4608 cells. These unit modules
are placed in supermdules which are separate gas-tight enclosures. There
are two types of unit modules and super-modules, the difference being in the number of cells 
in the row and column, in one there are 48 rows and 96 columns while in the 
other there are 96 rows and 48 columns. 
There are 4 supermodules in each
plane, arranged in such a way that each  forms
 one quadrant of the detector. The detector can be
opened as two halves around the beam pipe.

The design of 
signal processing electronics for the PMD is based on that of the 
tracking chambers
of dimuon spectrometer in ALICE.
The signals from the cells are processed using MANAS chips which provide analog multiplxed
output for 16 channels.
The readout of the analog multiplexed signals is performed,
after digitization, using another custom
built chip MARC and the CROCUS system.

The PMD will be mounted at 360cm from the center of the TPC in ALICE and will
nominally cover the pseudorapidity region 2.3-3.5 with full azimuthal
acceptance.

\begin{table}
\begin{center}
\caption{Parameters of STAR and ALICE PMDs}
\label{pmd-collider}
\vskip 4mm
\begin{tabular} {|l|l|l|} \hline
Parameter & STAR@RHIC & ALICE@LHC \\ \hline
($\eta,\phi$) coverage &2.3-3.8, 2$\pi$ & 2.3-3.5, 2$\pi$ \\
Number of planes & Preshower, veto & preshower, veto \\
Number of super-modules & 24 & 8 \\
Number of unit modules & 144 & 48 \\
Number of cells & 83K & 220K \\
Distance from vertex & 540~cm & 360~cm \\
Cell cross-section & 1 cm$^2$ & 0.23 cm$^2$\\
Cell depth & 8 mm & 5 mm \\ \hline
\end{tabular}
\end{center}
\end{table}

\subsection{Photon reconstruction}

The photon reconstruction 
is achieved by clustering the hits and then applying 
suitable algorithm to reject charged hadron hits. For the WA93/WA98 PMDs, 
where only the preshower plane of the detector was used, a simple algorithm 
based on a threshold on the cluster signal was used for the 
discrimination~\cite{wa98-eta}. A more detailed algorithm based on
 neural network technique has also been investigated for PMD 
with charged particle veto~\cite{nn} and used in the study of
physics performance of ALICE PMD~\cite{alice-pmd-tdr}.

Due to interactions
of charged hadrons in lead converter and the nature of the discrimination
algorithm, the set of clusters finally labeled as 'photons' have some
contaminants. The efficiency of photon reconstruction is usually around 70\%
and the purity of the photons in the accepted sample is also around 70\%.
These numbers vary with the system, centrality and pseudorapidity as
studied using event generators and Monte-Carlo methods.

\section{Physics results}

The measurement of photon distributions in the forward rapidity region
has made significant contributions to the physics at the SPS energy.
The lower p$_T$ cutoff for the PMD is only about 30~MeV/c~\cite{wa93-eta,alice-pmd-tdr}, 
thus enabling it to
detect a large fraction of the photon spectrum. The presence of a charged
particle multiplicity detector (SPMD) overlapping in phase space with the PMD 
has helped to greatly enhance the scope of physics studies using the PMD.

\subsection{Pseudorapidity and multiplicity distributions}

Pseudorapidity distribution of photons has been measured at various centralities 
using the PMD in 
WA93 experiment for S+Au collisions at 200.A~GeV~\cite{wa93-eta}. 
In  general the shape of the distribution is
well reproduced 
by the event generators like VENUS~\cite{venus} 
but the absolute magnitude is somewhat underpredicted. Using the measurement of photon
multiplicity by the PMD and the transverse e.m. energy by the MIRAC calorimeter, it was
possible to study the centrality dependence of $<p_T>$ of photons~\cite{wa93-pt}.

\begin{figure}[h]
\begin{minipage}{18pc}
\includegraphics[width=18pc]{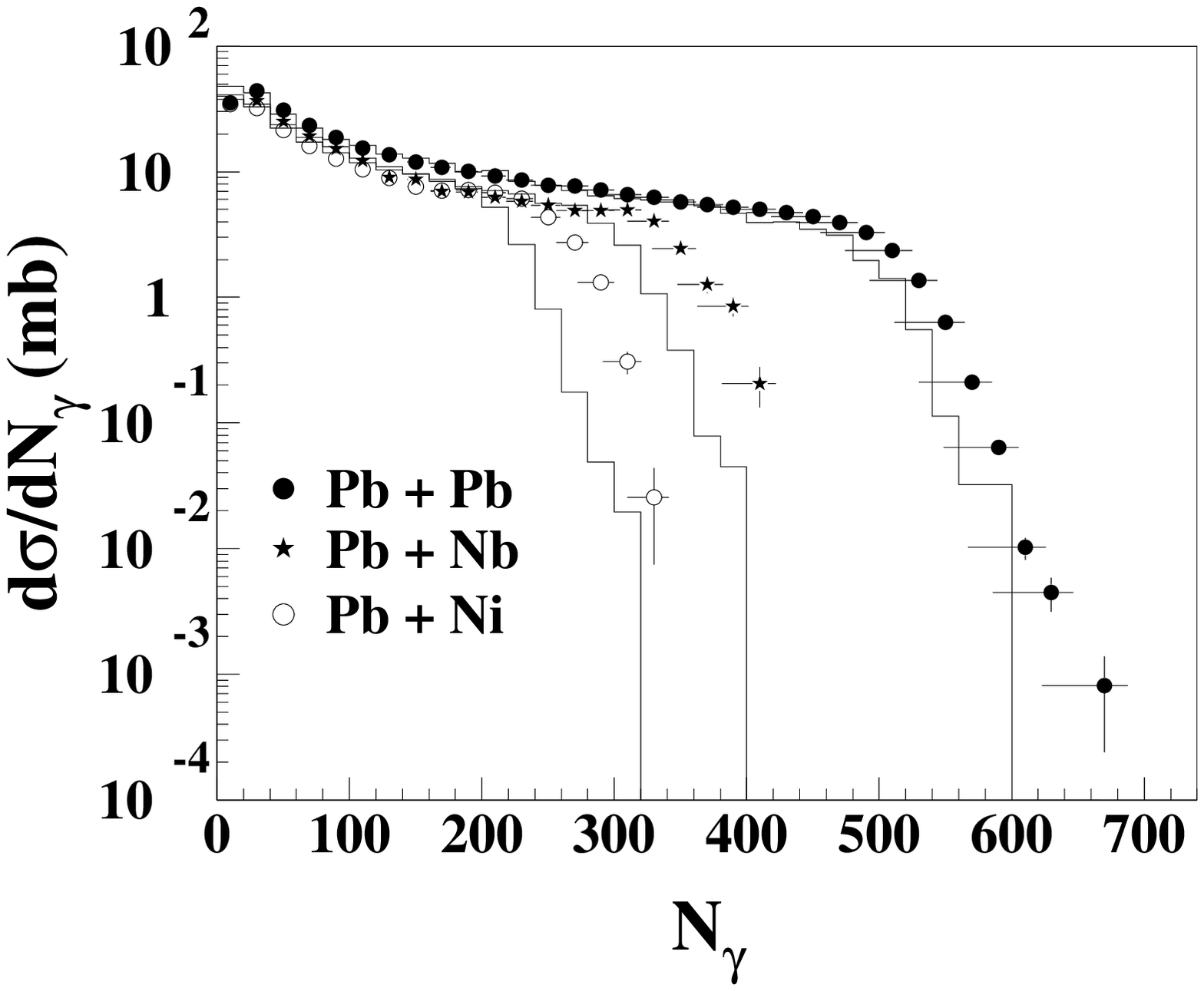}
\caption{\label{wa98-photon-multiplicity}
Minimum bias inclusive photon cross sections for Pb+Ni, Pb+Nb, 
and Pb+Pb reactions at 158$\cdot$AGeV. Solid histograms are the
corresponding distributions obtained from the VENUS event generator.}
\end{minipage}\hspace{2pc}%
\begin{minipage}{18pc}
\includegraphics[width=18pc]{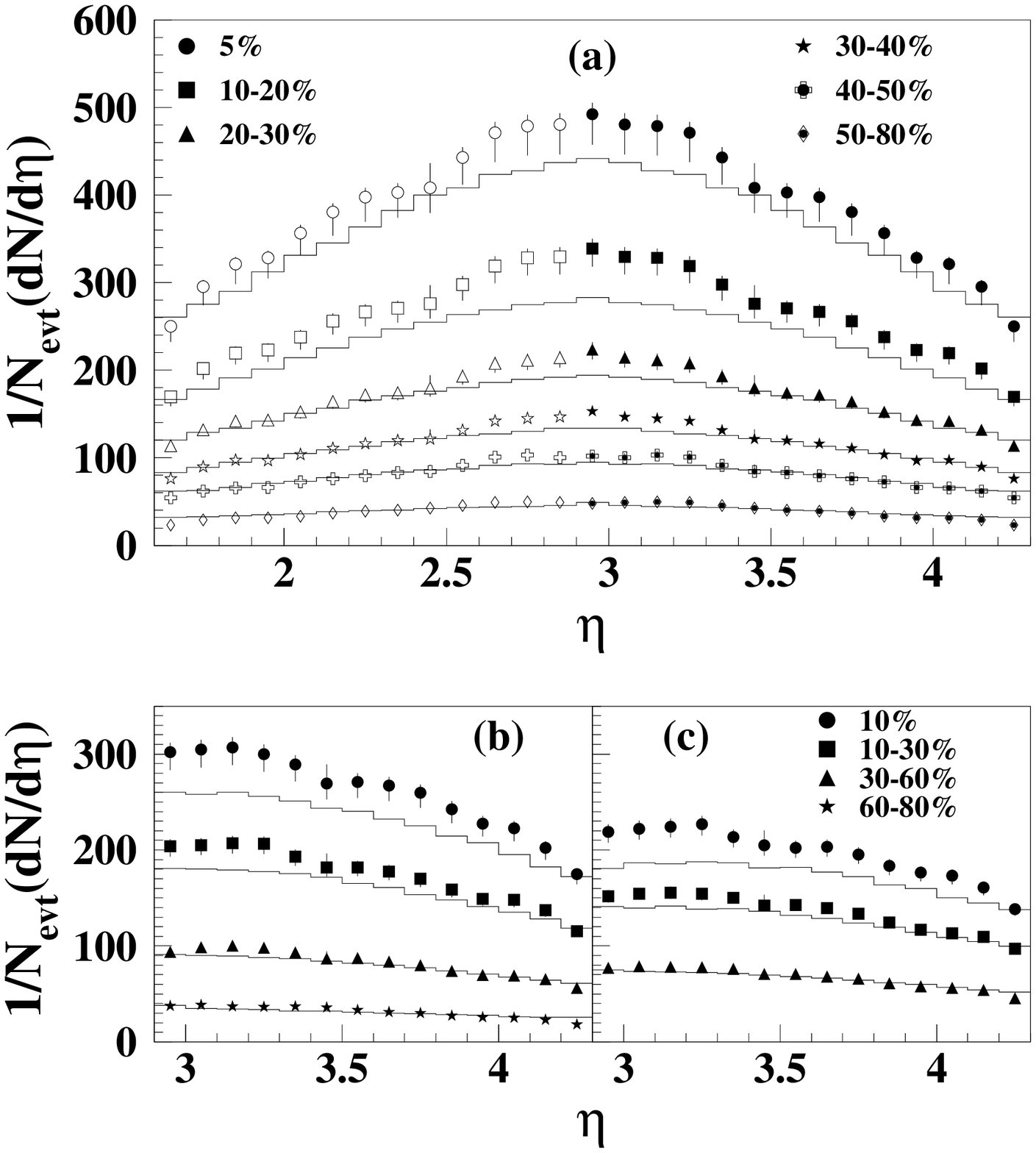}
\vspace*{-1.5cm}\caption{\label{wa98-photon-rapidity}
Pseudo-rapidity distributions of photons at various centralities in Pb induced 
reactions at 158$\cdot$AGeV on (a) Pb, (b) Nb and (c) Ni targets.
The solid histograms are the corresponding distributions
obtained from  the VENUS event generator.}
\end{minipage} 
\end{figure}

The WA98 PMD  measured photon production for  Pb+Pb, Pb+Nb and Pb+Ni systems
at 158.A~GeV projectile energy. A systematic study of these data showed
that VENUS again underpredicts the magnitude of pseudorapidity 
distributions~\cite{wa98-eta}.
Fig.~\ref{wa98-photon-multiplicity} shows the minimum bias inclusive photon
cross-section for the three systems along with the predictions of VENUS
event generator with default parameters. The shape of these distributions
is governed by the collision geometry. For asymmetric collisions of Nb and Ni
targets small shoulders are present around N$_\gamma$ of 300 and 200 
respectively. This shoulder is produced when a decrease in the impact 
parameter     leads to little increase in particle production and the 
cross-sections for these small impact parameter collisions pile up at a
fixed N$_\gamma$.

The pseudorapidity distributions for the three systems studied are shown
in Fig.~\ref{wa98-photon-rapidity} for various centralities. For the
symmetric Pb+Pb collisions filled symbols represent the measured data and
open symbols are reflections around $\eta_{cm}$ (=2.92). The histograms
show VENUS predictions. The discrepancy between data and VENUS is $\sim$10\%
for central collisions around mid-rapidity and decreases at larger impact
parameters. For asymmetric systems the discrepancy between data and VENUS
is comparatively larger.

An important observation of these systematic studies of photon production
over a wide range of collision systems and centralities has been the scaling
relation for total number of photons. It is found that this scales as a
power law dependence on the number of participants, N$_{part}$, the exponent
being 1.12$\pm$0.03~\cite{wa98-eta}.

First results on photon distribution in the forward region at 62.4~GeV energy
recently became available with the PMD in the STAR experiment~\cite{starphoton}. These 
are described in detail in the article by B. Mohanty et al. in these
proceedings~\cite{bedanga-talk}. At RHIC the HIJING model underpredicts photon  production as
measured by the PMD. However the AMPT model compares favorably well.

\begin{figure}[h]
\vspace*{-0.5cm}
\centerline{\includegraphics[scale=0.5]{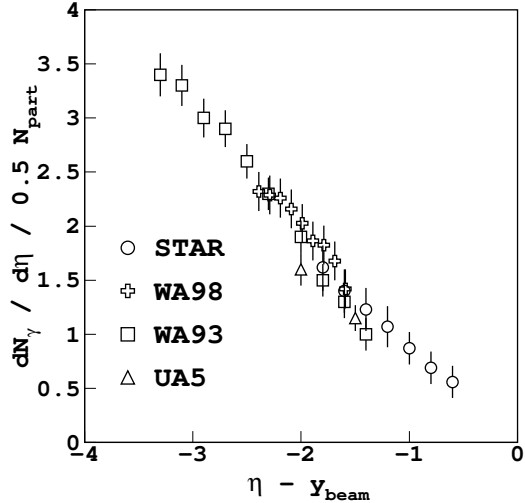}}
\vspace*{-0.5cm}
\caption{Limiting fragmentation as seen in the photon production.}
\label{lim-photon}
\end{figure}

In Fig.~\ref{lim-photon} we have plotted the photon 
pseudorapidity distribution per participant nucleon
 in Au+Au collisions at 62.4A~GeV as a function of $\eta$-y$_{beam}$ for 
central collisions. Also
superposed are the data from the WA98 experiment for the Pb+Pb collisions at
17.A~GeV c.m. energy, from the WA93 experiment for S+Au collisions at 20.A~GeV 
c.m. energy and from the
UA5 experiment for $pp$ collisions
 at 546 GeV. The data at SPS and at
RHIC are found to be consistent with each other, suggesting that photon
production follows limiting fragmentation behaviour. It is further found that
the data are consistent with centrality- and energy- independent beahaviour
similar to the results for identified charged particles~\cite{bedanga-talk}.

\subsection{Azimuthal anisotropy and flow}

In the search for evidence of the 
phase transition from hadronic matter to quark gluon plasma in relativistic
heavy-ion collisions, it is important to establish that thermalization occurs
in the reacting system. 
In  non-central collisions the overlap volume, 
when projected onto the transverse plane, is not azimuthally symmetric.
It has a smaller size along the direction of impact parameter
than in the perpendicular direction.  If thermal equilibrium
is reached, collective flow develops with a velocity
proportional to the pressure gradient,
which is larger along the direction of
impact parameter than along the perpendicular direction.
Matter is thus expected to flow preferentially
in the reaction plane 
which should result in an azimuthal anisotropy of the distribution
of particles.
In head-on collisions the  anisotropy 
should disappear, even if thermalization occurs,  
due to the azimuthal symmetry of the overlap volume. Thus the study of 
anisotropy as a function of centrality assumes significance in the study of
phase transitions.

\begin{figure}[h]
\begin{minipage}{18pc}
\includegraphics[width=18pc]{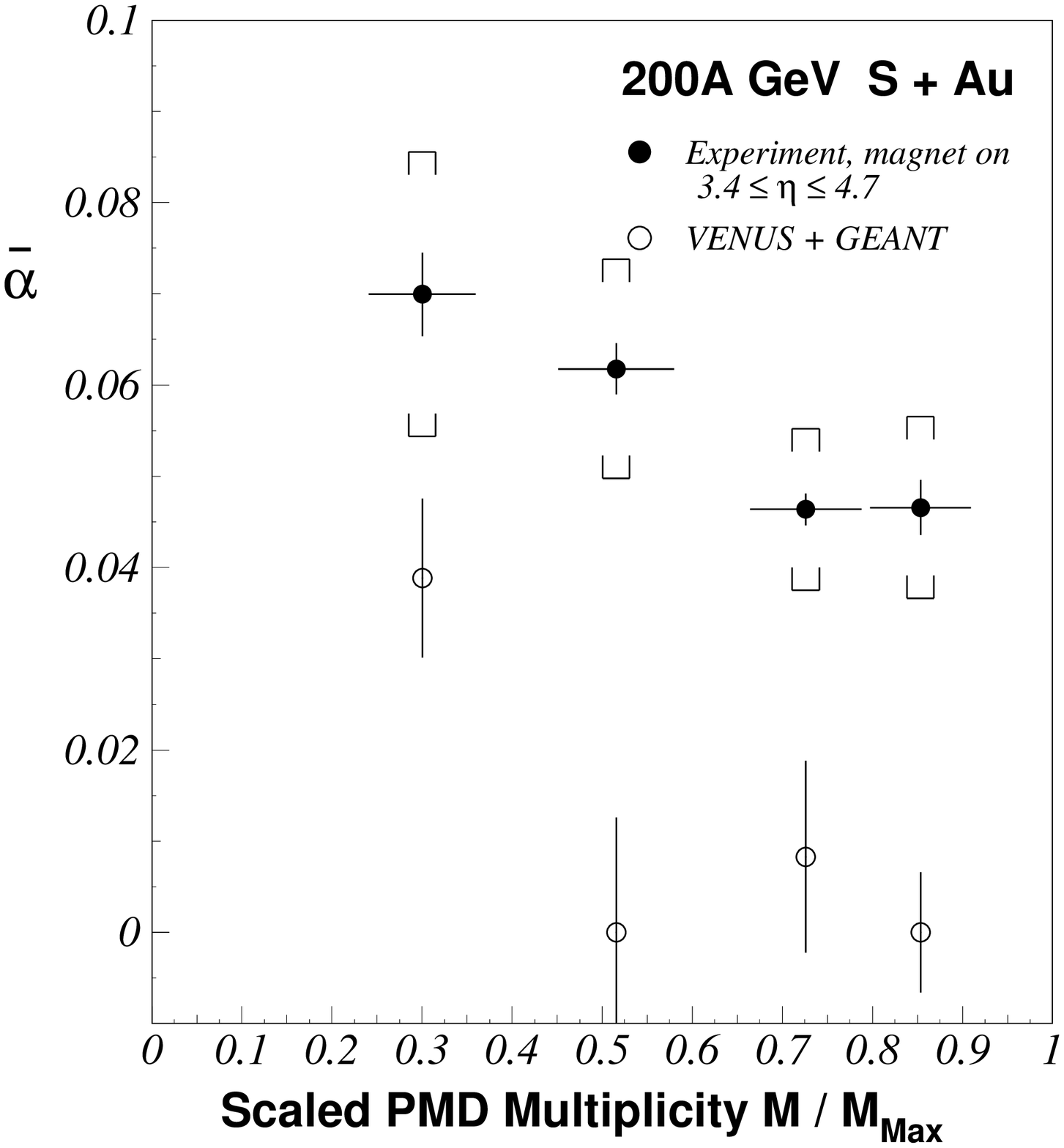}
\vspace*{-10mm}\caption{\label{wa93-flow}
The  true  anisotropy $\bar{\alpha}$ as  a function  of  the centrality 
(scaled  multiplicity)   
for  the  experimental  data  (filled circles)  and  
VENUS  +  GEANT  simulations  (open  circles).}  
\end{minipage}\hspace{2pc}%
\begin{minipage}{18pc}
\includegraphics[width=18pc]{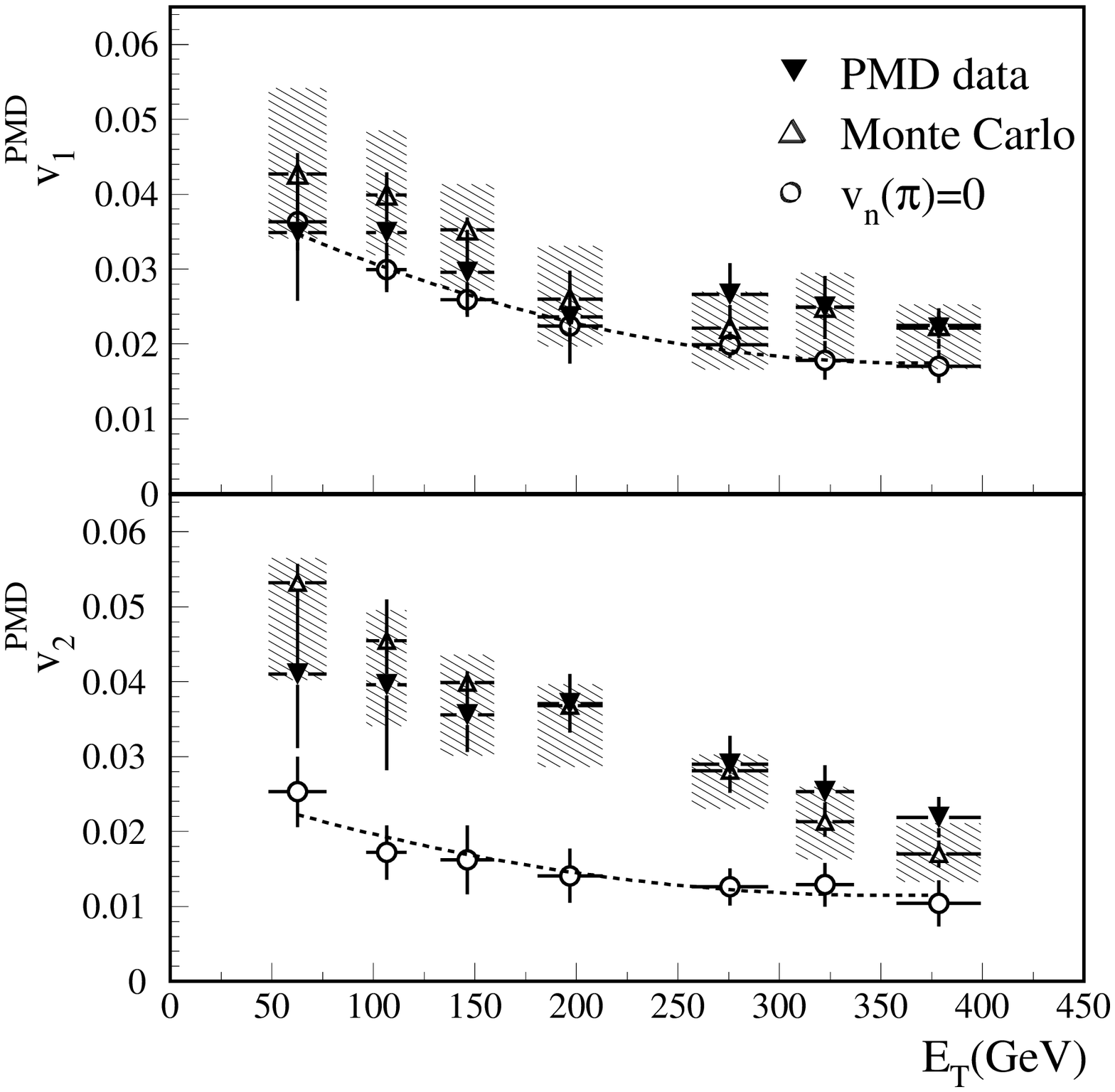}
\caption{\label{wa98-flow}
First order, $v_1$, and second order, $v_2$, anisotropy coefficients of the
azimuthal distributions of photon hits for different centralities
characterized by the measured transverse energy.
The boxes indicate the extent of the estimated systematic errors.}
\end{minipage} 
\end{figure}

{\it  The first observation of collective 
flow at the SPS energy came from the PMD data~\cite{wa93-flow}}.
The anisotropy parameter\footnote{At that time the 
nomenclature of various flow coefficients were not standardized. In Ref.
~\cite{wa93-flow} the word 'directed flow' has been used for what is now 
commonly 
understood as elliptic flow and the notation $\overline{\alpha}$ is actually
$v_2$.} $\overline{\alpha}$ 
 is plotted in Fig.~\ref{wa93-flow}
for the four centrality bins for the data (filled circles) 
and for the simulated data (open circles). The 
vertical  bars  indicate  statistical  errors  only.  The  horizontal  bars 
indicate  the  rms  multiplicity  of  the  bin.  The  estimated  systematic 
errors  are  indicated  by  the  bracket  on  each  data  point. 
The simulated events from VENUS event generator show a small anisotropy
similar in magnitude to the estimated systematic error. The non-zero values are
essentially due to $\pi^0$ decay correlations and finite multiplicity effects.

The extracted true anisotropies in the data are small but significantly greater
than the simulation which does not include flow effects. 
The observed
anisotropy decreases with decreasing impact parameter
from $\overline{\alpha} \approx 0.07$ for
peripheral collisions to   $\overline{\alpha} \approx 0.045$
for central collisions. This dependence is similar, but the values are
roughly half as large, as has been predicted by hydrodynamical model 
calculations \cite{yves92}.

The effect of photon correlation due to $\pi^0$ decay and other detector
related effects on the deduced anisotropy values from photon measurements
have been investigated in detail~\cite{raniwala}. It has been shown that
the measured anisotropy in photons can be used to deduce the anisotropy in 
parent neutral pions and a scaling relation between these two has been
extracted in terms of an experimentally measurable quantity.

This concept has been tested and established in the analysis of data from the
WA98 experiment~\cite{wa98-flow}.
For the study of anisotropy using data from the WA98 PMD, all the 
decay and detector
effects were considered within a Monte-Carlo framework using the measured
flow values of charged particles in the same experiment within an
overlapping part of phase space and assuming that $\pi^0$ flow value is the
same as that for charged particles (which are predominately charged pions).
The results are shown in Fig.~\ref{wa98-flow} for both the PMD data and the
simulation using input $\pi^0$ anisotropy. The photon anisotropy coefficients
extracted from the simulated PMD data are consistent within errors with the 
measured PMD results. 

The use of PMD data to determine the event plane gives a strong $\pi^0$ decay 
effect. This is shown by open circles in Fig.~\ref{wa98-flow} which show 
simulation results of anisotropy in PMD for isotropic $\pi^0$ emission. 
It is seen that
the first order anisotropy coefficients are dominated by the decay effect. 
Although second order coefficients also have significant decay contribution,
the measured values are much higher. 

This is the first time that a fully consistent result is obtained both from
charged particle anisotropy studies and photon anisotropy studies which are
in agreement with each other.

\subsection{Chiral symmetry restoration and disoriented chiral condensates}

The restoration of chiral symmetry and its subsequent breaking through a phase
 transition has been predicted  to create Disoriented Chiral Condensates
(DCC)~\cite{bedanga-review}. 
This phenomenon has been predicted to cause anomalous fluctuations in
the relative production of charged and neutral pions in high energy hadronic 
and nuclear collisions in certain phase space regions (domains). 
It is predicted that the neutral pion fraction
$f = \frac{N_{\pi^o}}{N_{\pi^o}+N_{\pi^{\pm}}}$
should follow a 1/$\sqrt f$ distribution in the case of DCC formation
in contrast to the binomial 
distribution for generic particle production.
A number of methods has been proposed to study the formation of DCC (see references in
\cite{bedanga-review}).

The formation of DCC has been studied extensively using the data 
from the two detectors, SPMD and PMD,
measuring respectively the multiplicity of charged particles and photons
 in overlapping part of phase space in the WA98 experiment.

\subsubsection{Correlation and wavelet analysis :}

DCC has been studied in the WA98 experiment by various methods, both in the
full overlapping phase space of the two detectors and also within smaller
azimuthal domains~\cite{wa98-global-dcc,wa98-local-dcc,wa98-central-dcc}.
Approaches based on direct correlation between charged particle and photon
multiplicities event-by-event and discrete wavelet transform method have been
used to study the distributions and their widths (rms values) looking for
evidence of non-statistical fluctuation when compared with various mixed 
event formulations.

By constructing a set of mixed events to study various detector effects, the
analysis showed that there exist non-statistical fluctuation 
individually in photon and
charged particle production. However a correlated fluctuation has not been
observed by this method of analysis. These results have been further compared
with the predictions of a simple DCC model where DCC effect is introduced in
the particle distribution of an event from a VENUS event generator at freezout
by pairwise flipping of
charged and neutral pions
according to the 1/$\sqrt f$ probability in a selected
($\eta-\phi)$ domain. The resulting distribution,
after allowing $\pi^0$ to decay,
 is then analysed using the
same correlation and wavelet analysis methods.
An upper limit on the
production of DCC-like fluctuation is estimated to be 10$^{-2}$ for 
$\Delta\phi$ between 45-90$^\circ$ and 3$\times$10$^{-3}$ for $\Delta\phi$ 
between 90-135$^\circ$ for the top 5\% centrality.

\begin{figure}[h]
\begin{minipage}{18pc}
\includegraphics[width=18pc]{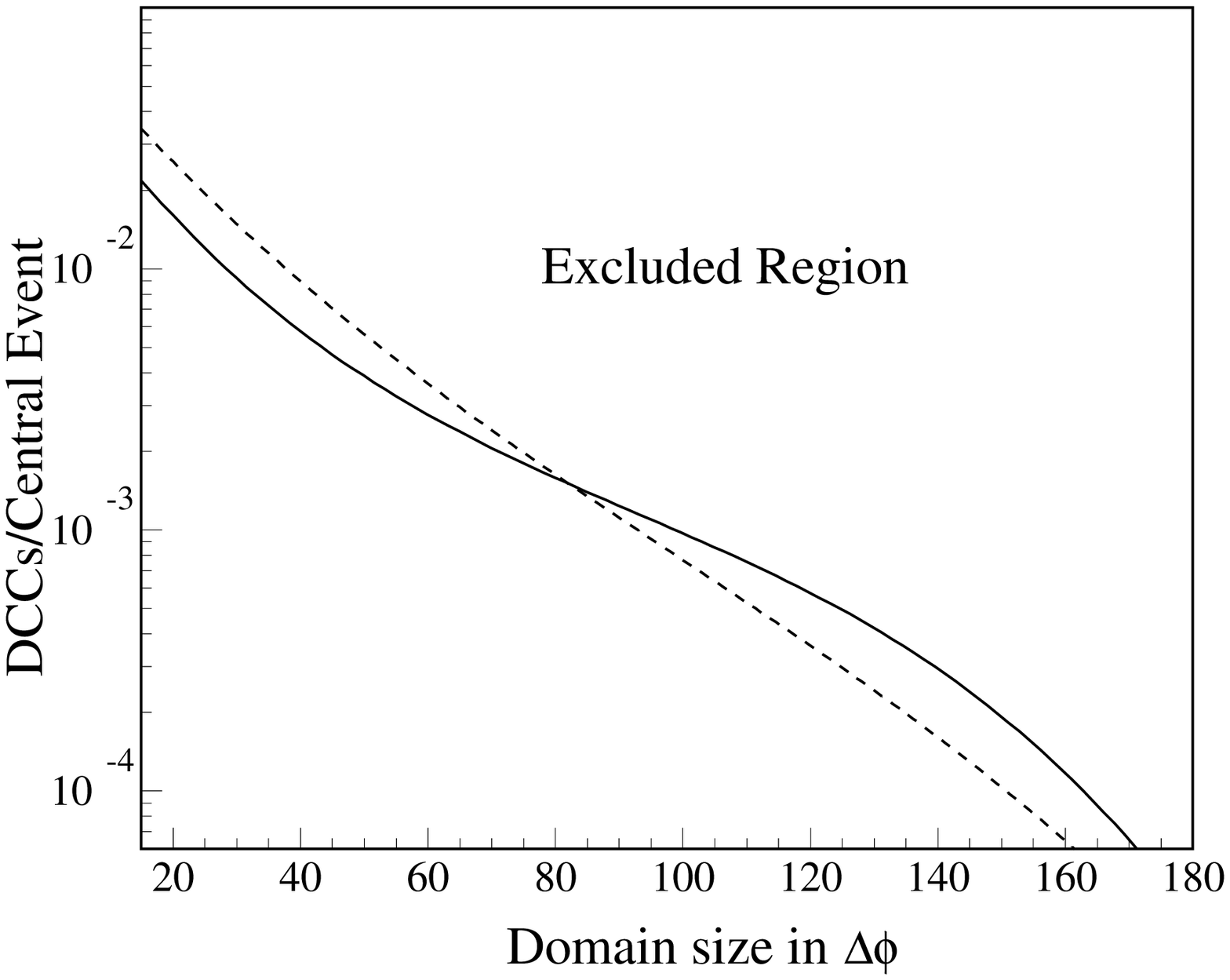}
\vspace*{-5mm}\caption{\label{dcc-upperlimit}The 90\% confidence level
upper limit on DCC formation for central Pb+Pb collisions at 158A~GeV,
 as a function of domain size in azimuthal angle within the context of a simple
DCC model and the measured photon and charged particle multiplicities
in the interval 2.9$\le\eta\le$3.75.}
\end{minipage}\hspace{2pc}%
\begin{minipage}{18pc}
\includegraphics[width=18pc]{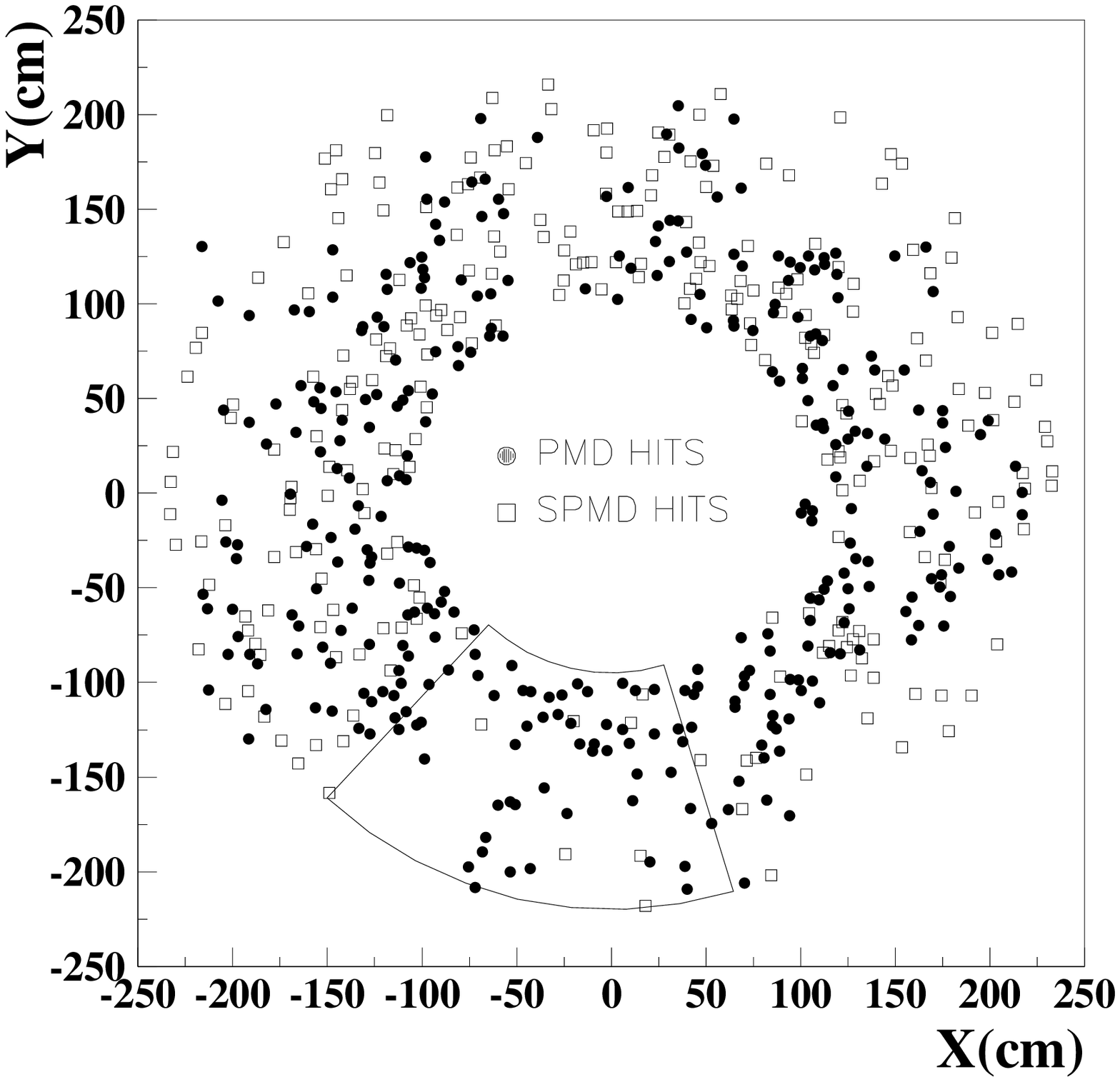}
\caption{\label{x-y-plot}
 Plot showing photon hits (PMD) and charged particle hits (SPMD)
  in an azimuthal plane. The marked  patch  shows a
   region where number of charged particles are very less compared to photons.}
\end{minipage} 
\end{figure}

Further studies have also been carried out for the effect of 
centrality~\cite{wa98-central-dcc}. The 
upper limit curves are shown in Fig.~\ref{dcc-upperlimit}
as a function of the azimuthal window for two centralities, top 5\% (solid 
lines) and top 5-10\% (dashed lines). It is observed that the fluctuations may
be more at lower centralities.

\subsubsection{Sliding window method :}

Several methods 
have been 
developed~\cite{arun,mma-swm} 
for the study of unusual structures in events arising due to
non-statistical fluctuations resulting
from phase transitions. The sliding window method (SWM) is particularly appealing for
the study of charge-neutral fluctuations and search for
Centauro and anti-Centauro like events through model-independent direct observation.
Some details of the method and preliminary results for the WA98 experiment
are given in Ref.~\cite{mma-swm}.
A typical event with an anti-Centauro like patch identified by the 
SWM is shown in Fig.~\ref{x-y-plot}, 
 displaying PMD hits (filled circles) 
and SPMD hits (open boxes), projected onto the PMD plane,
 within the overlapping ($\eta,\phi$) zone.
 The marked 60$^{\circ}$ patch 
corresponds to $f$ = 0.8, this has only 7 charged particle hits
as compared to 55 photon hits.



The percentage of events having 
exotic patches in data, with $f$-values beyond 4.5$\sigma$ of the generic $f$-
distribution,
 is found to be 0.39$\pm$0.016, 
as compared to 0.081 $\pm$0.007 in mixed events and 0.013 $\pm$0.008 in GEANT 
simulated VENUS events.
These patches are almost uniformly distributed in azimuth. 
Although a direct comparison with the results of 
upper limits, described in the previous section is not possible, 
the present value of 0.39\% seems within the expected range.
 The present results are based on direct
observation of event structure and are model-independent. The results, however, are 
preliminary and further checks are in 
progress to study the detector related effects as this is an important
observation.

\subsection{Multiplicity fluctuation}

\begin{figure}
\begin{minipage}{18pc}
\includegraphics[width=18pc]{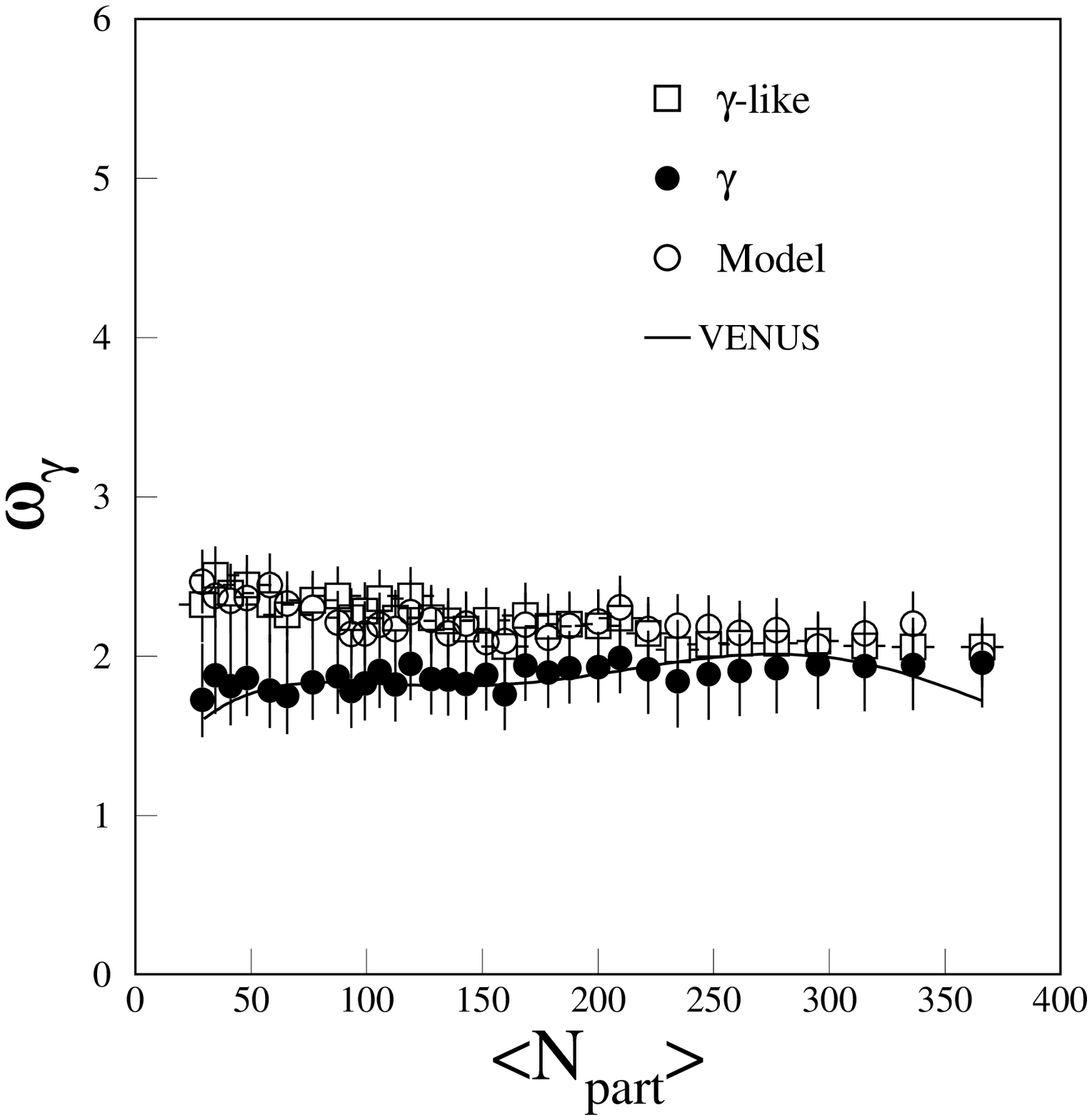}
\caption{\label{wa98_fluc_photon}
The relative fluctuations, $\omega_{\gamma}$ of photons
as a function of number of participants. These are compared to
calculations from a participant model and those from VENUS event generator.}
\end{minipage}\hspace*{2pc}
\begin{minipage}{18pc}
\includegraphics[width=18pc]{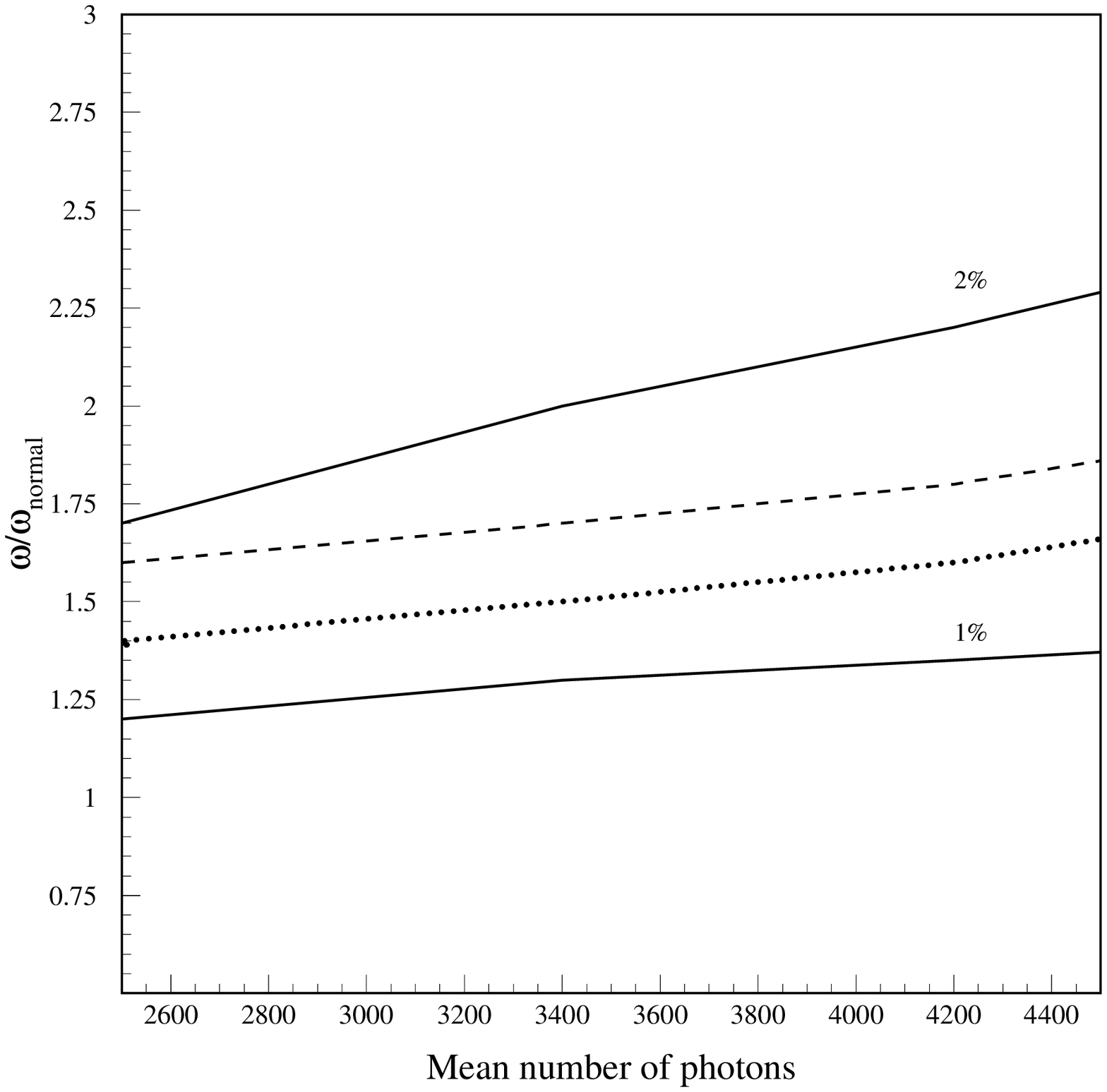}
\caption{\label{atdr.4.7}
Fluctuation $\omega / \omega_{normal}$ as a function of centrality.
The dotted and dashed lines show the limits due to detector effects for a
 standalone PMD  and for PMD with upstream material in the ALICE environment. }
\end{minipage}
\end{figure}

  With the production of large number of particles on an event-by-event basis
  in high energy heavy ion collisions, it is very important to study anomalous
  fluctuations of multiplicity which is a major signal of QGP phase
  transition \cite{heiselberg}. It has been proposed that event-by-event
  multiplicity fluctuations can arise from density fluctuations or droplet
  formation due to first order phase transition \cite{baym}. Large
  fluctuations in multiplicity are also expected at the tricritical point of
  QCD phase diagram \cite{stephanov}. 
Event-by-event  fluctuations in the multiplicities of
 photons and charged particles have been studied in the WA98 experiment at
  CERN-SPS~\cite{wa98_fluc}. Fig.~\ref{wa98_fluc_photon} shows the 
relative  fluctuations, $\omega_{\gamma}$ (= $\sigma^{2}/N_{\gamma}$) 
of photons as a function of mean 
number of participants. The data presented show the
fluctuations in $\gamma-like$ clusters (open squares) and photons after correction for 
efficiency and purity (filled circles). 

These are compared to calculations from a participant model~\cite{participant} and those 
from VENUS event generator. The centrality dependence of  charged particle
multiplicity fluctuation in the measured data is found to agree reasonably
well with those obtained from the participant model. However, for photons the
multiplicity fluctuation has been found to be lower compared to those
obtained from the participant model.

\section{Future possibilities}

The study of anisotropy of produced particles will remain a major goal of the
PMD. One can deduce the flow values for neutral pions and compare with those
for charged pions measured using other subsystems. 
 The PMD will also provide the event plane information for the study of anisotropic
emission of particles in any other rapidity window, both in STAR and in ALICE
experiments. The usefulness of PMD has already been studied in the context of
anisotropic emission of J/$\Psi$ particles in the ALICE dimuon spcetroemter~\cite{jpsi}. 

The STAR experiment has a forward time projection chamber (FTPC) in a 
pseudorapidity region which is 
overlapping with the coverage of the PMD. The FTPC provides a measure of
charged particle multiplicity as well as their momenta. Using the momentum
selection of charged particles, the study of DCC-like charged neutral 
fluctuation should be possible  with improved results as has been shown by
simulation studies~\cite{poslim}. It is shown that the p$_T$ 
selection of charged particles enhances the DCC signal considerably more
than that of photons.

In the ALICE experiment at the LHC, the much higher multiplicity of produced 
particles should result in highly reduced statistical fluctuation and 
consequently enhanced sensitivity to dynamical fluctuation. Possible physics
studies with the PMD in ALICE have been presented in 
Refs.~\cite{alice-pmd-tdr,alice-atdr}. The forward multiplicity detector (FMD)
in ALICE has the ($\eta,\phi$) coverage overlapping with those of the PMD.
The  FMD-PMD pair is expected to be used for the study of charged-neutral
fluctuation at the LHC energy. 

The sensitivity of the PMD to small fluctuations has been studied using the
fluctuation measure $\omega$ = $\sigma_{\gamma}^{2}/N_\gamma$ in a fast
simulation technique where dynamical fluctuation has been introduced in the
particle distribution to make it wider. The resulting distribution is again
analyzed for the same fluctuation measure. Fig.~\ref{atdr.4.7} shows the
ratio of fluctuation measure after and before introducing the dynamical 
component plotted as a function of the mean multiplicity of photons in PMD.
It is
found that dynamical fluctuation as low as 2\% can be observed with the PMD
even at lower multiplicities. As the statistical fluctuation decreases with
increasing multiplicity,  the
difference between the limits set by detector
effects and the dynamical fluctuation increases, providing more sensitivity
at higher multiplicity.

\section{Summary}

  Using a fine granularity preshower photon multiplicity detector 
  (PMD), 
important contribution has ben made  to the
study of ultra-relativistic heavy ion collisions. Beginning with a detector
made of plastic scintillator pads and wavelength shifting optical fibres for
the SPS experiments, the PMD for the RHIC and LHC colliders has been
made in the form of honeycomb gas proportional counters using a 
novel concept of extended cathode. Among the important
studies at the SPS are
 first observation of collective flow and a detailed study
of localized charged-neutral fluctuation as a probe for DCC. 
In addition the PMD's contribution include study of
  pseudo-rapidity distributions of photons, scaling of photon
  multiplicity with number of participating nucleons, 
  centrality dependence of $<p_{T}>$ of photons and
  event-by-event fluctuations in photon multiplicity.
The PMD is taking data at the RHIC collider and is expected to be 
installed in ALICE experiment at the LHC in 2007. The RHIC program 
has already paid dividends in the study of limiting fragmentation 
of identified particles. The study of DCC will be continued at the RHIC using 
the PMD and the FTPC in STAR experiment and at the LHC using 
the PMD and the FMD in the ALICE experiment.
At the LHC, larger multiplicity will be helpful 
in studying the event-by-event fluctuation in global observables with better precision.

\ack

The experimental program for the study of quark-gluon plasma at CERN and 
BNL 
is funded by the Department of Atomic Energy and the Department of 
Science and
 Technology of the Government of India. 
In addition the program has also derived 
financial support from the University Grants Commission and  the Council of
Scientific and Industrial Research of the Government of India, 
the STAR Collaboration, the CERN NMS grants,
and the Indo-German
 Exchange Program in various
 stages of the work. We acknowledge the collaborative work of all 
the team members, past and present,
of the WA93 collaboration, WA98 collaboration, STAR collaboration and ALICE 
Collaboration. We acknowledge the  extensive 
support received 
from the engineering and technical manpower and computer facility personnel
at BNL and CERN and also 
 at all the collaborating institutions, namely VECC Kolkata, 
IOP Bhubaneswar, Panjab University Chandigarh,
University of Rajasthan Jaipur, University of Jammu, Jammu and GSI
 Darmstadt.

\end{document}